\documentclass[aps,preprint,tightenlines,]{revtex4}

\usepackage{amssymb}
\usepackage{amsmath}
\usepackage{graphicx}

\begin{document}

\title{Comparison of quantum mechanical and classical trajectory
calculations of cross sections for ion-atom impact ionization of negative -
and positive -ions for heavy ion fusion applications }
\author{Igor D. Kaganovich, Edward A. Startsev and Ronald C. Davidson}
\affiliation{Plasma Physics Laboratory, Princeton University, Princeton, NJ 08543}
\date{\today}

\begin{abstract}
Stripping cross sections in nitrogen have been calculated using the
classical trajectory approximation and the Born approximation of quantum
mechanics for the outer shell electrons of 3.2GeV I$^{-}$ and Cs$^{+}$ ions.
A large difference in cross section, up to a factor of six, calculated in
quantum mechanics and classical mechanics, has been obtained. Because at
such high velocities the Born approximation is well validated, the classical
trajectory approach fails to correctly predict the stripping cross sections
at high energies for electron orbitals with low ionization potential.
\end{abstract}

\maketitle

\section{Introduction}

Ion-atom ionizing collisions play an important role in many applications,
such as heavy ion inertial fusion \cite{HIF reference}, collisional and
radiative processes in the Earth's upper atmosphere \cite{atmosphere},
ion-beam lifetimes in accelerators \cite{accelerators life time}, atomic
spectroscopy \cite{spectroscopy} and ion stopping in matter \cite{beam
stopping}, and are also of considerable academic interest in atomic physics
\cite{Review atomic physics}.

To estimate the ionization and stripping rates of fast ions propagating
through gas or plasma, the values of ion-atom ionization cross sections are
necessary. In contrast to the electron \cite{Voronov} and proton \cite{Rudd,
Rudd 2} ionization cross sections, where experimental data or theoretical
calculations exist for practically any ion and atom, the knowledge of
ionization cross sections by fast complex ions and atoms is far from
complete \cite{Shvelko book}. While specific values of the cross sections
for various pairs of projectile ions and target atoms have been measured at
several energies \cite{our PoP hif, Olson exp, Watson exp}, the scaling of
cross sections with energy and target or projectile nucleus charge has not
been experimentally mapped.

There are several theoretical approaches to cross section calculations.
These include: classical calculations that make use of a classical
trajectory and the atomic electron velocity distribution functions given by
quantum mechanics [this approach is frequently referred to as the classical
trajectory Monte Carlo (CTMC) approach]; quantum mechanical calculations
based on the Born, eikonal or quasiclassical approximations, and so forth
\cite{Shvelko book}. All approaches are computationally intensive, and the
error and range of validity have to be assessed carefully before making any
approximations or applying the results.

Classical trajectory calculations are simpler to perform in comparison with
quantum mechanical calculations. Moreover, in some cases the CTMC
calculations yield results close to the quantum mechanical calculations \cite%
{our PoP hif, Mueller new, Our new}. The reason for similar
results lies in the fact that the Rutherford scattering cross
section is identical in both classical and quantum mechanical
derivations \cite{Landau book}. Therefore, when an ionizing
collision is predominantly a consequence of electron scattering at
small impact parameters close to the nucleus, the quantum
mechanical uncertainty in the scattering angle is small compared
with the angle itself, and the classical calculation can yield an
accurate description \cite{Bohr, my PAC Xsection}. But this is not
always a case, as we demonstrate below. For fast projectile
velocities and low ionization potentials, the difference between
the classical and quantum mechanical calculations of ionization
cross section can be as large as a factor of six for parameters to
relevant to heavy ion fusion cross sections.

In the present analysis, we consider at first only the stripping
cross section of loosely bound electron orbitals of $I^{-}$ and
$Cs^{+}$ ions colliding with a neutral atom of nitrogen, or with a
fully stripped nitrogen ion with $Z_{T}=7$ (for comparison).
Atomic units are used throughout this paper with $e=\hbar
=m_{e}=1$, which corresponds to length normalized to $a_{0}=\hbar
^{2}/(m_{e}e^{2})=0.529\cdot 10^{-8}cm,$ velocity normalized to $%
v_{0}=e^{2}/\hbar =2.19\cdot 10^{8}cm/s$, and energy normalized to $%
E_{0}=m_{e}v_{0}^{2}=2Ry=27.2eV$, where $Ry$ is the Rydberg energy. The
normalizing coefficients are retained in all equations for robust
application of the formulas. For efficient manipulation of the formulas, it
is worth noting that the normalized velocity is $v/v_{0}=0.2\sqrt{E[keV/amu]}
$, where $E$ is energy per nucleon in $keV/amu$. Therefore, $25keV/amu$
corresponds to the atomic velocity scale.

The typical scale for the electron orbital velocity with ionization
potential $I_{nl}$ is $v_{nl}=v_{0}\sqrt{2I_{nl}/E_{0}}$. Here, $n,l$ is the
standard notation for the main quantum number and the orbital angular
momentum quantum number \cite{Landau book}. The collision dynamics is very
different depending on whether $v$ is smaller or larger than $v_{nl}$.

\section{Behavior of cross sections at large values of projectile velocity $%
v>v_{nl}$}

When $v>>v_{nl}$, the projectile interaction with the target atom occurs for
a very short time, and the interaction time decreases as the velocity
increases. For $3.2GeV$ $I^{-}$ ions, envisioned for heavy ion fusion
applications, the projectile velocity in atomic units is $32v_{0}$, while
the electron orbital velocity is $v_{nl}=0.5v_{0}$ for the first ($3.06eV$)
ionization potential of $I^{-}$, and $v_{nl}=1.3v_{0}$ for the first ($%
22.4eV $) ionization potential of $Cs^{+}$. Therefore, we shall use the
limit $v>>v_{nl}$.

In the limit, where $v>v_{0}Z_{T}$ and $v>>v_{nl}$, the Born approximation
of quantum mechanics can be used \cite{Landau book, Mueller new}. The first
inequality assures that the nitrogen atomic potential can be taken into
account as a small perturbation (the Born approximation); the second
inequality allows us to use the unperturbed atomic wave function.

\begin{figure}[tbp]
\includegraphics{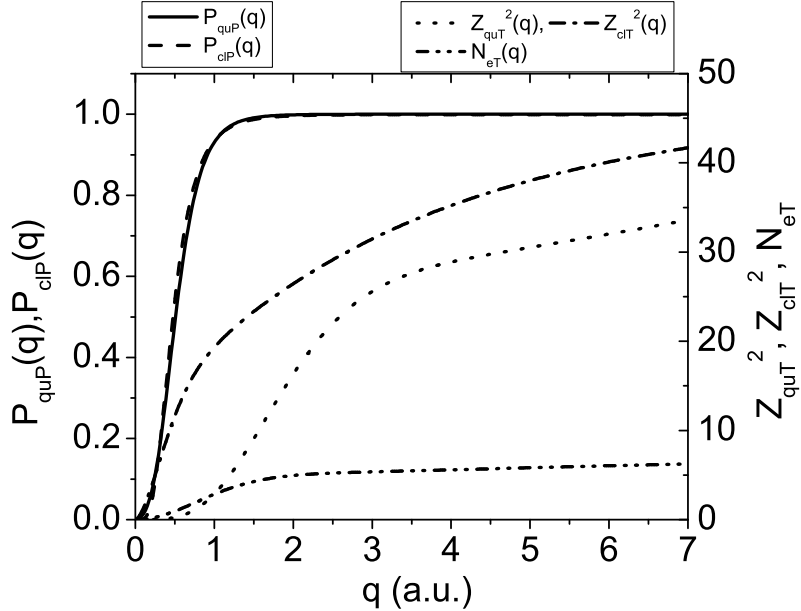}
\caption{Shown in the figure is a comparison of the ionization probabilities
[$P_{quP}(q)$ in Eq.(\protect\ref{P quantum}), and $P_{clP}(q)$ in Eq.(%
\protect\ref{P classical mechanics})] and the effective charges [$Z_{quT}(q)$
$N_{eT}(q)$ in Eq.(\protect\ref{Z quantum}), and $Z_{clT}(q)$ in Eq.(\protect
\ref{Z cl effective})] in quantum and classical mechanics for $3.2GeV$ $I^{-}
$ ions colliding with a nitrogen atom. Ionization of only the outer electron
shell is considered (here, $I_{nlP}=3eV$). }
\label{Fig1}
\end{figure}

In both classical mechanics and in the Born approximation, the ionization
cross section can be recast in the form \cite{Shvelko book, Bethe, Bethe
book, Our new},%
\begin{equation}
\sigma =\int_{0}^{\infty }P_{P}(q)\frac{d\sigma }{dq}dq,
\label{sigma ionization as integral}
\end{equation}%
where $P_{P}(q)$ is the probability of electron stripping
from the projectile when the electron acquires the momentum $q$, and $%
d\sigma /dq$ is the differential cross section for scattering with momentum $%
q$.

\begin{figure}[tbp]
\includegraphics{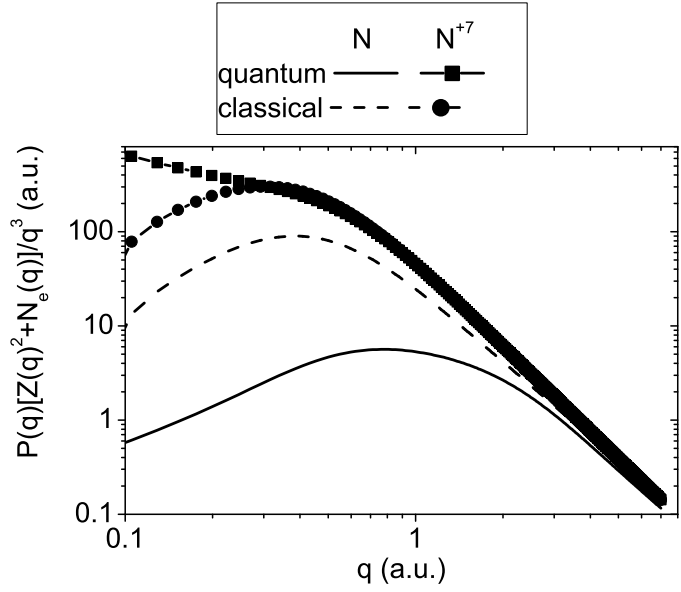}
\caption{Plots of differential cross sections for stripping of $I^{-}$ ions
by nitrogen atoms and fully stripped ions.}
\label{Fig.2}
\end{figure}

In quantum mechanics, $P_{quP}(q)$ can be expressed by the square of the
corresponding matrix element of transition from the initial state $|nl>$ to
the state of the ejected electron $|\mathbf{k}>$ with momentum $\mathbf{k}$,
integrated over all $\mathbf{k}$. This gives
\begin{equation}
P_{quP}(q)=\int \left\vert <nl|e^{i\mathbf{q\cdot r}}|\mathbf{k}>\right\vert
^{2}d^{3}\mathbf{k.}  \label{P quantum}
\end{equation}%
The analytical form of $P_{quP}(q)$ for hydrogen-like electron functions is
given in Ref. \cite{Bethe}. In classical mechanics, $P_{clP}(q)$ is given by
the integral over the electron velocity distribution function $f(\mathbf{v}%
_{e}\mathbf{)}$ defined by%
\begin{equation}
P_{clP}(q)=\int \Theta \left( \mathbf{q\cdot v}_{e}+\frac{q^{2}}{2m_{e}}%
-I_{nl}\right) f(\mathbf{v}_{e}\mathbf{)dv}_{e}\mathbf{.}
\label{P classical mechanics}
\end{equation}%
Classical mechanics prescribes the electron velocity distribution function
(EVDF) for hydrogen-like orbitals as a microcanonical ensemble, where
\begin{equation*}
f\left( \mathbf{v}_{e}\right) =Cv_{e}^{2}\int \delta \left( \frac{%
m_{e}v_{e}^{2}}{2}-\frac{e^{2}Z_{T}}{r}+I_{nl}\right) r^{2}dr.
\end{equation*}%
Here, $C$ is a normalization constant defined so that $\int \,f\left(
v_{e}\right) dv_{e}=1$, and $\delta (...)$ denotes the Dirac delta-function.
Interestingly, the EVDF for a hydrogen-like electron orbitals is identical
in both the quantum mechanical and classical calculations \cite{Landau book}%
, with
\begin{equation}
\,f\left( v_{e}\right) \,=\frac{32v_{nl}^{7}}{\pi }\frac{v_{e}^{2}}{\left[
v_{e}^{2}+v_{nl}^{2}\right] ^{4}},
\end{equation}%
where $v_{nl}$ is the scale of the electron orbital velocity defined by
\begin{equation}
v_{nl}=v_{0}\sqrt{2I_{nl}/E_{0}}.
\end{equation}%
In the Born approximation of quantum mechanics, $d\sigma /dq$ is
given by \cite{Landau book, Shevelko paper}
\begin{equation}
\frac{d\sigma }{dq}=8\pi a_{0}^{2}\frac{v_{0}^{2}(m_{e}v_{0})^{2}}{v^{2}}%
\frac{Z_{quT}^{2}(q)+N_{eT}(q)}{q^{3}},  \label{d sigma dq qm}
\end{equation}%
where
\begin{equation}
Z_{quT}(q)=\left\vert Z_{T}-\sum_{nl}F_{nlT}(q)\right\vert
,\;N_{eT}(q)=[N_{eT}^{total}-\sum_{nlT}\left\vert F_{nlT}(q)\right\vert
^{2}].  \label{Z quantum}
\end{equation}%
Here, $Z_{quT}(q)$ is the effective charge, subscript $qu$ stands for
quantum mechanics, $F_{nlT}(q)=\int e^{i\mathbf{q\cdot r}}\rho _{nlT}(r)d^{3}%
\mathbf{r}$ is the form factor of the target atom's orbital $nl$ with the
electron density $\rho _{nlT}(r)$, and $N_{eT}^{total}$ is the total number
of electrons in the target atom [$N_{eT}(q\rightarrow \infty )=$ $%
N_{eT}^{total}$].

In classical mechanics, $d\sigma /dq$ is given by
\begin{equation}
\frac{d\sigma }{dq}=2\pi \rho \frac{d\rho }{dq}.  \label{d sigma d ro}
\end{equation}%
Here, $\rho (q)$ is the impact parameter for a collision resulting in the
momentum transfer $q.$ For fast collisions, $q$ is mainly perpendicular to
the projectile velocity, and $q$ is determined by integration of the
electric field of the target atom on the electron, which gives%
\begin{equation}
q(\rho )=-\frac{2\rho }{v}\int_{\rho }^{\infty }\frac{dU_{T}}{dr}\frac{1}{%
\sqrt{r^{2}-\rho ^{2}}}dr,  \label{q(r)}
\end{equation}%
where $U_{T}(r)$ is the atomic potential of the target atom. To compare the
classical calculation with the quantum mechanical calculation, we recast
Eqs.(\ref{d sigma d ro}) and (\ref{q(r)}) into a form similar to Eq.(\ref{d
sigma dq qm}), introducing the effective charge $Z_{clT}(q)$ defined by%
\begin{equation}
Z_{clT}(q)=\frac{qv}{2m_{e}a_{0}v_{0}^{2}}\sqrt{-q\rho (q)\frac{d\rho }{dq}},
\label{Z cl effective}
\end{equation}%
where subscript $cl$ stands for classical mechanics. Note that for the bare
target ion, $U_{T}=-e^{2}Z_{T}/r$ and $Z_{clT}(q)=Z_{T}$. Finally, making
use of the effective charge in Eq.(\ref{Z cl effective}), the differential
cross section in classical mechanics takes on a form similar to Eq.(\ref{d
sigma dq qm}) in quantum mechanics, i.e.,
\begin{equation}
\frac{d\sigma }{dq}=8\pi a_{0}^{2}\frac{v_{0}^{2}(m_{e}v_{0})^{2}}{v^{2}}%
\frac{Z_{clT}(q)^{2}+N_{eT}^{total}}{q^{3}}.
\end{equation}%
Here, the final term accounts for ionization by the $N_{eT}^{total}$ target
electrons.

Figure 1 shows a comparison of the ionization probabilities [$P_{quP}(q)$ in
Eq.(\ref{P quantum}), and $P_{clP}(q)$ in Eq.(\ref{P classical mechanics})]
and the effective charges [$Z_{quT}(q)$ in Eq.(\ref{Z quantum}), and $%
Z_{clT}(q)$ in Eq.(\ref{Z cl effective})] in quantum mechanics and classical
mechanics for $3.2GeV$ $I^{-}$ ions colliding with a nitrogen atom.
Ionization of only the outer electron shell is considered (here, $%
I_{nlP}=3.06eV$, approximating as a hydrogen-like orbital).

Figure 2 shows that for stripping by neutral atoms, the main contributions
arise from intermediate momenta in the range $q=0.5-1$, while for  stripping
by the bare target nucleus, small values of $q$ make the largest
contribution to the cross section, which corresponds to large impact
parameters (due to the Coulomb long-range interaction). Because $%
P_{quP}>P_{clP}$ for $q<<1$, but $Z_{quT}<Z_{clT}(q)$, the quantum
mechanical cross sections are larger than the classical  stripping cross
sections for  stripping by the bare nucleus, but smaller than the classical
stripping cross sections for the atoms. Carrying out the integration in Eq. (%
\ref{sigma ionization as integral}) gives the  stripping cross sections for
only one electron from the outer electron shell for different ions with the
same velocity $v=32v_{0}$ colliding with a nitrogen atom. The results are
shown in Table 1 for $3.2GeV$ $I^{-}$ ions; in Table 2 for $3.35GeV$ $Cs^{+}$
ions; and in Table 3 for $25MeV$ $H^{-}$.\newline

\begin{tabular}{lll}
$\sigma ,10^{-16}cm^{2}$ & quantum & classical \\
N & 0.08 & 0.47 \\
N$^{+7}$ & 2.5 & 1.29%
\end{tabular}%
\newline

\textbf{Table 1.} Cross section for stripping of $3.2GeV$ $I^{-}$ ions
colliding with a nitrogen atom and a fully stripped nitrogen ion ( stripping
of only one electron from the outer electron shell is considered here with $%
I_{nlP}=3.06eV$ ).\newline

\begin{tabular}{lll}
$\sigma ,10^{-16}cm^{2}$ & quantum & classical \\
N & 0.045 & 0.10 \\
N$^{+7}$ & 0.32 & 0.17%
\end{tabular}%
\newline

\textbf{Table 2.} Cross section for stripping of $3.35GeV$ $Cs^{+}$ ions
(the same velocity as $3.2GeV$ $I^{-}$) colliding with a nitrogen atom or a
fully stripped nitrogen ion ( stripping of only one electron from the outer
electron shell is considered here with $I_{nlP}=22.4eV$ ). \newline

\begin{tabular}{lll}
$\sigma ,10^{-16}cm^{2}$ & quantum & classical \\
N & 0.10 & 1.34 \\
N$^{+7}$ & 12.5 & 5.05%
\end{tabular}%
\newline

\textbf{Table 3.} Cross section for stripping of $25MeV$ $H^{-}$
ions (the same velocity as $3.2GeV$ $I^{-}$) colliding with a
nitrogen atom or a fully stripped nitrogen ion ( stripping of only
one electron from the outer electron shell is considered here with
$I_{nlP}=0.75eV$ ). \newline \newline
Figure 3 shows the same
results as in Fig.2, but the results are obtained for $3.35GeV$
$Cs^{+}$ ions (ionization of only one outer electron shell is
considered here with $I_{nlP}=22.4eV$ ). Note that $3.35GeV$
$Cs^{+}$ is chosen to have the same velocity as a $3.2GeV$ $I^{-}$
ion.

In the limit $v>>v_{nl},$ the stripping cross section by a fully stripped
ion can be analytically evaluated. The Bohr formula, derived by means of
classical mechanics, neglects the electron atomic velocity, and gives for
the cross section \cite{Bohr}
\begin{equation}
\sigma ^{Bohr}(v,I_{nl},Z_{p})=2\pi Z_{p}^{2}a_{0}^{2}\,\,\frac{%
v_{0}^{2}E_{0}}{v^{2}I_{nl}}.  \label{Bohr}
\end{equation}%
Accounting for the electron atomic velocity gives an additional factor of $%
5/3$ \cite{Our new}. The Bethe formula \cite{Bethe} derived by means of the
Born approximation of quantum mechanics gives
\begin{equation}
\sigma ^{Bethe}=\sigma ^{Bohr}(v,I_{nl},Z_{p})\left[ 0.566\ln \left( \frac{v%
}{v_{nl}}\right) +1.261\right] .  \label{Bethe}
\end{equation}%
The results of cross sections calculations using Eq.(\ref{Bohr}) with a
factor $5/3$ and the result in Eq.(\ref{Bethe}) coincide with the results in
Tables 1, 2 and 3 of stripping cross sections by a fully stripped nitrogen
ions calculated in classical trajectory approximation and the Born
approximation of quantum mechanics, respectively.

The stripping cross sections calculated in classical trajectory
approximation for $Cs^{+}$ and $I^{-}$ ions by fully stripped nitrogen ions
is only factor 2-3 larger than the stripping cross sections by neutral
nitrogen atoms, which is in qualitative agreement with the observations in
Ref.\cite{Olson exp}. However, there is a large difference, up to a factor
30, in the stripping cross sections calculated in the Born approximation of
quantum mechanics.

It is evident that the stripping of $Cs^{+}$ ions by fully stripped nitrogen
ions decreases by a factor of $22.4eV/3eV=7.5$ compared with $I^{-}$ ions,
which is in agreement with the Bohr [Eq.(\ref{Bohr})] and Bethe [Eq.(\ref%
{Bethe})] formulas. The stripping cross sections for $Cs^{+}$ and $I^{-}$%
ions by neutral nitrogen atoms differ by only a factor of 2. In classical
mechanics, because the interaction potential is a strong function of the
separation, to transfer a considerably larger momentum requires a rather
small decrease in impact parameter. This is why, notwithstanding the large
difference in ionization potential by a factor of $7$, the difference
between the two cross sections is only a factor of 2. Table 3 shows that the
difference between the quantum and classical treatments increases for
smaller ionization potentials (compare Table 3 with Table 1).

\begin{figure}[tbp]
\includegraphics{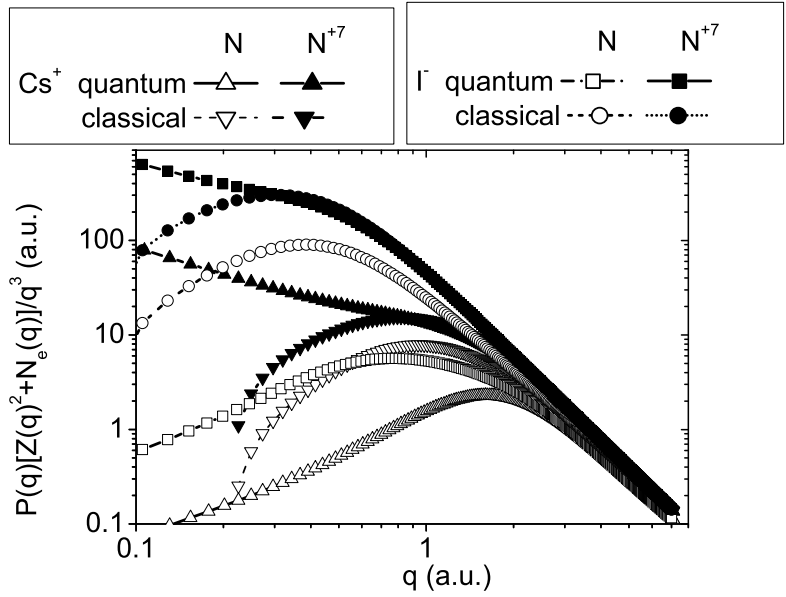}
\caption{Plots of the differential cross sections of ionization for $Cs^{+}$
and $I^{-}$ ions by nitrogen atoms and fully stripped ions.}
\label{Fig3}
\end{figure}

The reason for such a large difference between the quantum mechanical and
classical mechanical stripping cross sections for $I^{-}$ can be easily
understood from the example of elastic electron scattering from the shielded
Coulomb potential $U(r)=\exp (-r/a_{0})/r$. The differential cross section
for elastic scattering is shown in Fig.4 .
\begin{figure}[tbp]
\includegraphics{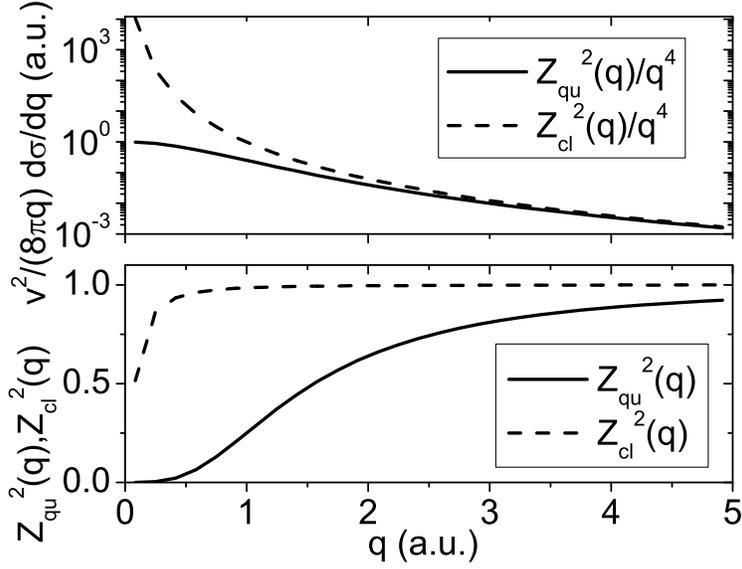}
\caption{Plots of the differential cross sections for the shielded Coulomb
potential for $v=32v_{0}$.}
\label{fig4}
\end{figure}
For the shielded Coulomb potential, direct application of the Born
approximation gives \cite{Landau book}%
\begin{equation}
\frac{d\sigma }{qdq}=8\pi a_{0}^{2}\frac{v_{0}^{2}(m_{e}v_{0})^{2}}{v^{2}}%
\frac{1}{(q^{2}+m_{e}^{2}\hbar ^{2}/a_{0}^{2})^{2}},
\end{equation}%
and the total cross section is $\sigma =4\pi a_{0}^{2}v_{0}^{2}/v^{2}.$ The
total classical cross section, obtained from integrating $\int \rho d\rho $,
diverges because of the contributions from large $\rho $ (small $q)$.
Evidently, the quantum mechanical cross section departs from the Rutherford
scattering formula for $q/(m_{e}v_{0})<1$, whereas the classical mechanical
cross section departs from the Rutherford scattering formula only for $%
q/(m_{e}v_{0})<2v_{0}/v$ [see Eq.(\ref{q(r)}) and Fig.4]. Therefore, the
classical differential cross section differs from the quantum mechanical
result by a factor of $[v/(2v_{0})]^{4}$, which for $v=32v_{0}$ gives a
difference in small-angle differential cross section of up to a factor of $%
10^{4}$ (see Fig.4).

Tables 4 and 5 are similar to Tables 1 and 2, but the calculations are
carried out for ion energies 30 times smaller, in the range of $100MeV.$
Table 5 shows that the predictions of the classical and quantum mechanical
theories are similar for 100MeV ions. However, they are a factor two
different for $I^{-}$ ions, and the cross sections are the same within 10\%
accuracy for $Cs^{+}$ ions. The contribution from small $q$ to the stripping
cross section by a neutral nitrogen atom is smaller for $Cs^{+}$ ions than
for $I^{-}$ ions, thereby significantly reducing the stripping cross section
of $Cs^{+}$ ions compared with $I^{-}$ ions, especially for the calculation
in the classical trajectory approximation (see Tables 4 and 5, and Fig.5).%
\newline

\begin{tabular}{lll}
$\sigma ,10^{-16}cm^{2}$ & quantum & classical \\
N & 2.47 & 6.8 \\
N$^{+7}$ & 61 & 37%
\end{tabular}%
\newline

\textbf{Table 4.} Cross section for the stripping of $105MeV$ $I^{-}$ ions ($%
v=5.75v_{0}$) colliding with a nitrogen atom and a fully stripped nitrogen
ion (stripping of only one electron from the outer electron shell is
considered here with $I_{nlP}=3eV$ ).\newline

\begin{tabular}{lll}
$\sigma ,10^{-16}cm^{2}$ & quantum & classical \\
N & 1.36 & 1.4 \\
N$^{+7}$ & 6.6 & 5.2%
\end{tabular}%
\newline

\textbf{Table 5. }Cross section for the stripping of $110MeV$ $Cs^{+}$ ions (%
$v=5.75v_{0}$) colliding with a nitrogen atom and a fully stripped nitrogen
ion (stripping of only one electron from the outer electron shell is
considered here with $I_{nlP}=22.4eV$).

\begin{figure}[tbp]
\includegraphics{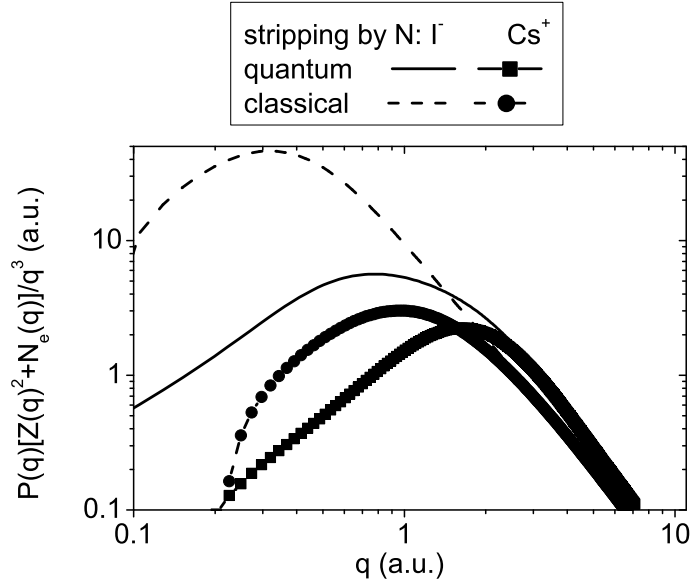}
\caption{Plots of the differential cross sections for stripping of 100MeV$%
Cs^{+}$ and 105MeV $I^{-}$ ions ($v=7.5v_{0}$) by nitrogen atoms.}
\label{Fig.5}
\end{figure}

\section{Calculation of total cross sections}

The total stripping cross section is defined as
\begin{equation}
\sigma ^{total}=\sum_{m}m\sigma _{m},
\end{equation}%
where $\sigma _{m}$ is the cross section for stripping $m$
electrons in each collision. This cross section is convenient to
use for electron production calculations. The stripping cross
section for any degree of ionization is defined as
\begin{equation}
\sigma =\sum_{m}\sigma _{m},
\end{equation}%
which is a convenient expression to use to determine the ion confinement
time in an accelerator. In the limit $v>>v_{nl}$, the calculation of the
total stripping cross section can be performed assuming that the stripping
from different electron orbitals occurs independently \cite{Shvelko book},
i.e.,%
\begin{equation}
\sigma ^{total}=\sum_{nl}N_{nl}\sigma _{nl,}  \label{summ of cross sections}
\end{equation}%
where $\sigma _{nl}$ is the stripping cross section  of only one
electron from the electron orbital $nl$, and $N_{nl}$ is the
number of electrons in the orbital. The structure of the electron
orbitals for $I^{-}$ ions is shown in Table 6.\newline

\begin{tabular}{lllllllllll}
$nl$ & 5p & 5s & 4d & 4p & 4s & 3d & 3p & 3s & 2p & 2s \\
N$_{nl}$ & 6 & 2 & 10 & 6 & 2 & 10 & 6 & 2 & 6 & 2 \\
I$_{nl}$ & 3.08 & 13.2 & 50.1 & 125.0 & 185.83 & 623.26 & 892.5 & 1.07e3 &
4.65e3 & 5.2e3 \\
$\sigma _{nl}$($v=32v_{0})$ & 0.080 & 0.054 & 0.030 & 0.018 & 0.013 & 5.5e-3
& 4.2e-3 & 3.6e-3 & 8.3e-4 & 7.3e-4 \\
$\sigma _{nl}$($v=5.75v_{0})$ & 2.45 & 1.65 & 0.92 & 0.52 & 0.39 & 0.12 &
0.078 & 0.062 & 5.8e-3 & 4.6e-3%
\end{tabular}%
\newline

\textbf{Table 6.} The structure of electron orbitals for $I^{-}$ ions and
the individual cross sections avaluated for an orbital electron in units of $%
10^{-16}cm^{2}$. \newline

Here, $nl$ denotes the atomic orbital quantum numbers, I$_{nl}$ is the
ionization potential in eV, and $\sigma _{nl}$ denotes the individual cross
section for an orbital electron in units of $10^{-16}cm^{2}$. The sum over
all orbitals gives $\sigma ^{total}=1.1\cdot 10^{-16}cm^{2}$ for 3.2GeV $%
I^{-}$ ions. To correctly account for multiple ionization, the inclusion of
multi-electron effects is necessary. This will be addressed in a future
publication. However, it is clear that the stripping cross section for any
degree of ionization by neutral atoms is limited by the geometrical cross
section of the atom (the geometrical cross section of a nitrogen atom is
much smaller than the geometrical cross section of a $Cs^{+}$ ion or a $I^{-}
$ion \cite{Periodic table}). The nitrogen atom geometric cross section is $%
\sigma _{N}=1.5\cdot 10^{-16}cm^{2}$\cite{Periodic table}, and therefore $%
\sigma <\sigma _{N}$ is expected. Preliminary estimates suggest that single
electron stripping is expected under these conditions.

For 105MeV $I^{-}$ ions, however, the sum over all orbitals gives
$\sigma ^{total}=33\cdot 10^{-16}cm^{2}$, whereas $\sigma
_{N}=1.5\cdot 10^{-16}cm^{2}.$ This indicates that multi-electron
ionization is expected. However, it is clear that the stripping
cross section for any degree of ionization is limited from above
by $\sigma _{N}=1.5\cdot 10^{-16}cm^{2}$.

The structure of the electron orbitals for $Cs^{+}$ ions, and the
individual cross sections for an orbital electron in units of
$10^{-16}cm^{2}$ are illustrated in Table 7. Note that a $Cs^{+}$
ion has the same number of electrons on each orbital as a $I^{-}$
ion.\newline

\begin{tabular}{lllllllllll}
$nl$ & 5p & 5s & 4d & 4p & 4s & 3d & 3p & 3s & 2p & 2s \\
N$_{nl}$ & 6 & 2 & 10 & 6 & 2 & 10 & 6 & 2 & 6 & 2 \\
I$_{nl}$ & 22.4 & 34.0 & 88.3 & 176 & 242 & 742 & 1.03e3 & 1.2e3 & 5.1e3 &
5.7e3 \\
$\sigma _{nl}$($v=32v_{0})$ & 0.044 & 0.037 & 0.022 & 0.014 & 0.011 & 4.8e-3
& 3.7e-3 & 3.2e-3 & 7.4e-4 & 6.5e-4 \\
$\sigma _{nl}$($v=5.75v_{0})$ & 1.35 & 1.12 & 0.66 & 0.41 & 0.32 & 0.098 &
0.065 & 0.052 & 4.7e-3 & 3.8e-3%
\end{tabular}%
\newline

\textbf{Table 7}. The structure of electron orbitals for $Cs^{+}$ ions and
the individual cross sections for an orbital electron in units of $%
10^{-16}cm^{2}$. \newline

For 3.35GeV $Cs^{+}$ ions colliding with a nitrogen atom with velocity $%
v=32v_{0}$ ($25MeV/amu$), the summation in Eq.(\ref{summ of cross sections})
over all orbitals gives $\sigma ^{total}=0.72\cdot 10^{-16}cm^{2}.$ This
estimate of the cross section is consistent with Olson's result in Ref.\cite%
{Olson exp}, $\sigma =2\cdot 10^{-16}cm^{2}$ for $25MeV/amu$ $Xe^{+}.$ Note
that the factor of three difference between the results presented in Table 7
and the results in Ref.\cite{Olson exp} is due to the fact that the cross
sections in Table 7 are predicted by making use of quantum mechanics,
whereas results in Ref.\cite{Olson exp} are classical trajectory
calculations, not applicable at such high projectile velocities.

For $110MeV$ $Cs^{+}$ ions colliding with a nitrogen atom, $v=5.75v_{0}$ ($%
0.8Mev/amu$) and the summation over all orbitals in Eq.(\ref{summ of cross
sections}) gives $\sigma ^{total}=21\cdot 10^{-16}cm^{2},$ whereas the
geometrical cross section of a nitrogen atom is only $\sigma _{N}=1.5\cdot
10^{-16}cm^{2}<<\sigma ^{total}.$ This indicates that multi-electron
ionization is expected, similar to $I^{-}$ ions at the same velocity. As
noted earlier, to correctly account for multiple ionization, multi-electron
calculations are necessary. However, it is clear that the stripping cross
section $\sigma $ for any degree ionization is limited by $\sigma
_{N}=1.5\cdot 10^{-16}cm^{2}$. This estimate of the cross section is
consistent with Olson's result \cite{Olson exp}, $\sigma ^{total}=4\cdot
10^{-16}cm^{2}$ for $2MeV/amu$ $Xe^{+}$. The inequality $\sigma
^{total}>\sigma _{N}$ indicates the important effect of multi-electron
events.

\section{\textbf{Conclusions}}

For low ionization potential, where a small momentum transfer $q$
contributes to stripping, the classical approach is not valid. For $3.2GeV$ $%
I^{-}$ ions, the classical trajectory approach overestimates by a factor of
six the stripping cross section by atomic nitrogen, and by a factor of two
the stripping cross section of $3.35GeV$ $Cs^{+}$ ions. For $110MeV$ $Cs^{+}$
ions and $105MeV$ $I^{-}$ ions colliding with a nitrogen atom at velocity $%
v=5.75v_{0}$ ($0.8Mev/amu$), multi-electron ionization is
expected. For a correct description of multiple ionization,
multi-electron calculations are necessary. However, it is clear
that the stripping cross section for any degree of ionization is
limited from above by the geometrical cross section of nitrogen,
with $\sigma _{N}=1.5\cdot 10^{-16}cm^{2}$, and should be be
similar in magnitude for $I^{-}$ ions and $Cs^{+}$ ions at
energies in the 100MeV range. (The geometrical cross section of a
nitrogen atom is much smaller than the geometrical cross section
of a $Cs^{+}$ ion or a $I^{-}$ion \cite{Periodic table}. This
effect is similar to the hole produced by a bullet piercing a
paper target, where the hole size is determined by the bullet
cross section, \textit{not} by the paper target.)

\textbf{Acknowledgments}

This research was supported by the U.S. Department of Energy. It is a
pleasure to acknowledge the benefits of useful discussion with Christine
Celata, Larry Grisham, Grant Logan and Art Molvik.


\begin{thebibliography}{99}
\bibitem{HIF reference} B. G. Logan, C. M. Celata, J. W. Kwan, E. P. Lee, M.
Leitner, P. A. Seidl, S. S. Yu, J. J. Barnard, A. Friedman, W. R. Meier, R.
C. Davidson, Laser and Particle Beams \textbf{20}, 369 (2002).

\bibitem{atmosphere} G. M. Keating, S. W. Bougher, J. Geophys. Res.- Space
Phys. \textbf{97 }(A4), 4189 (1992).

\bibitem{accelerators life time} H. Beyer, V. P. Shevelko (eds), \textit{%
Atomic physics with Heavy Ions} (Springer, Berlin 1999).

\bibitem{spectroscopy} A. Bogaerts, R. Gijbels, R. J. Carman, Spectrochimica
Acta Part B - Atomic Spectroscopy \textbf{53}, 1679 (1998).

\bibitem{beam stopping} C. Stockl, O. Boine-Frankenheim, M. Geissel, M.
Roth, H. Wetzler, W. Seelig, O. Iwase, P. Spiller, R. Bock, W. Suss, D. H.
H. Hoffmann, Nucl. Instrum. Meth. A \ \textbf{415}, 558 (1998).

\bibitem{Review atomic physics} S. Datz, G. W. F. Drake, T. F. Galagher, H.
Kleinpoppen, G. Zu Putlitz, Rev. Mod. Phys. \textbf{71}, S223 (1999).

\bibitem{Voronov} G. S. Voronov, Atomic Data and Nuclear Data Tables \textbf{%
65}, 1, (1997).

\bibitem{Rudd} M. E. Rudd, Y. -K. Kim, D. H. Madison, and J. W. Galallagher,
Rev. Mod. Phys. \textbf{64}, 441 (1992).

\bibitem{Rudd 2} M. E. Rudd, Y. -K. Kim, D. H. Madison, and T. J. Gay, Rev.
Mod. Phys. \textbf{57}, 965 (1985).

\bibitem{Shvelko book} R. K. Janev, L. P. Presnyakov, V. P. Shevelko,
\textit{Physics of Highly Charged Ions} (Springer, Berlin 1999).

\bibitem{our PoP hif} D. Mueller, L. Grisham, I. Kaganovich, R. L. Watson,
V. Horvat and K. E. Zaharakis, Physics of Plasmas, \textbf{8}, 1753 (2001).

\bibitem{Olson exp} R. E. Olson, R. L. Watson, V. Horvat, and K. E.
Zaharakis, J. Phys. B: At. Mol. Opt. Phys. \textbf{35}, 1893 (2002).

\bibitem{Watson exp} R. L. Watson, Y. Peng, V. Horvat, G. J. Kim, and R.E.
Olson, Phys. Rev.A \textbf{67}, 022706 (2003).

\bibitem{Mueller new} D. Mueller, L. Grisham, I. Kaganovich, R. L. Watson,
V. Horvat, K. E. Zaharakis and Y. Peng, Laser and Particle Beams \textbf{20}%
, 551 (2002).

\bibitem{Our new} I. D. Kaganovich, E. Startsev and R. C. Davidson,
"Ionization Cross-Sections in Ion-Atom Collisions for High Energy Ion
Beams", Proceedings of the 2003 Particle Accelerator Conference, in
preparation (2003).

\bibitem{Landau book} L. D. Landau and E. M. Lifshitz, \textit{Quantum
Mechanics} (Addison-Wesley Publishing Co., 1958).

\bibitem{Bohr} N. Bohr, K. Dan. Vidensk. Selsk. Mat.- Fys. Medd. \textbf{18}%
, N8 (1948).

\bibitem{my PAC Xsection} I. D. Kaganovich, E. Startsev and R. C. Davidson,
\textquotedblleft Evaluation of Ionization Cross Sections in Energetic
Ion-Atom Collisions,\textquotedblright\ Proceedings of the 2001 Particle
Accelerator Conference, (2001).
http://accelconf.web.cern.ch/AccelConf/p01/PAPERS/TPAH314.PDF

\bibitem{Bethe} H. Bethe, Ann. Phys. (Leipz.) \textbf{5}, 325 (1930).

\bibitem{Bethe book} H. A. Bethe and R. Jackiw, \textit{Intermidiate Quantum
Mechanics} (The Benjamin/Cummings Publishing Company, sec. ed., 1968).

\bibitem{Shevelko paper} V. P. Shevelko, I. Yu. Tolstikhina and Th.
Stoehlker, Nucl. Instr. Meth. B \textbf{184}, 295 (2001).

\bibitem{Periodic table} Periodic Table of the Elements in CRC Handbook of
Chemistry and Physics, 81st edition, 200-2001.
\end{thebibliography}
\end{document}